\documentclass[12pt]{article}

\usepackage{times}
\usepackage{graphicx}
\usepackage{amssymb}
\usepackage{amsthm}
\usepackage{amsmath}
\usepackage{dsfont}
\usepackage{bm}
\usepackage{mathrsfs}
\usepackage{bbold}
\usepackage{color}
\usepackage{xcolor}
\usepackage{soul}
\usepackage{hyperref}
\usepackage[superscript]{cite}
\usepackage[symbol]{footmisc}

\begin{document}
\title{Optical time travel: proposal for testing Hawking's Chronology Protection Conjecture in an optical analogue}
\author{David Bermudez$^1$ and Ulf Leonhardt$^{2*}$\\
$^1$Department of Physics,\\ Cinvestav, A.P. 14-740, 07000 Ciudad de M\'{e}xico, Mexico\\
$^2$Weizmann Institute of Science, Rehovot 761001, Israel
}
\maketitle
\begin{abstract}
Classical general relativity allows time travel, but Hawking [Phys. Rev. D {\bf 46}, 603 (1992)] conjectured that quantum mechanics ultimately prevents it. Here we propose a feasible experiment to test Hawking's Chronology Protection Conjecture in optics. In this scheme, time travel is inherently related to lasing, and the amplified vacuum fluctuations above laser threshold establish the quantum limits of time travel.\\

\noindent
Corresponding author: \href{mailto:ulf.leonhardt@weizmann.ac.il}{ulf.leonhardt@weizmann.ac.il}
\end{abstract}

\newpage

\section{Introduction}

\begin{figure}[t!]
	\begin{center}
		\includegraphics[width=18pc]{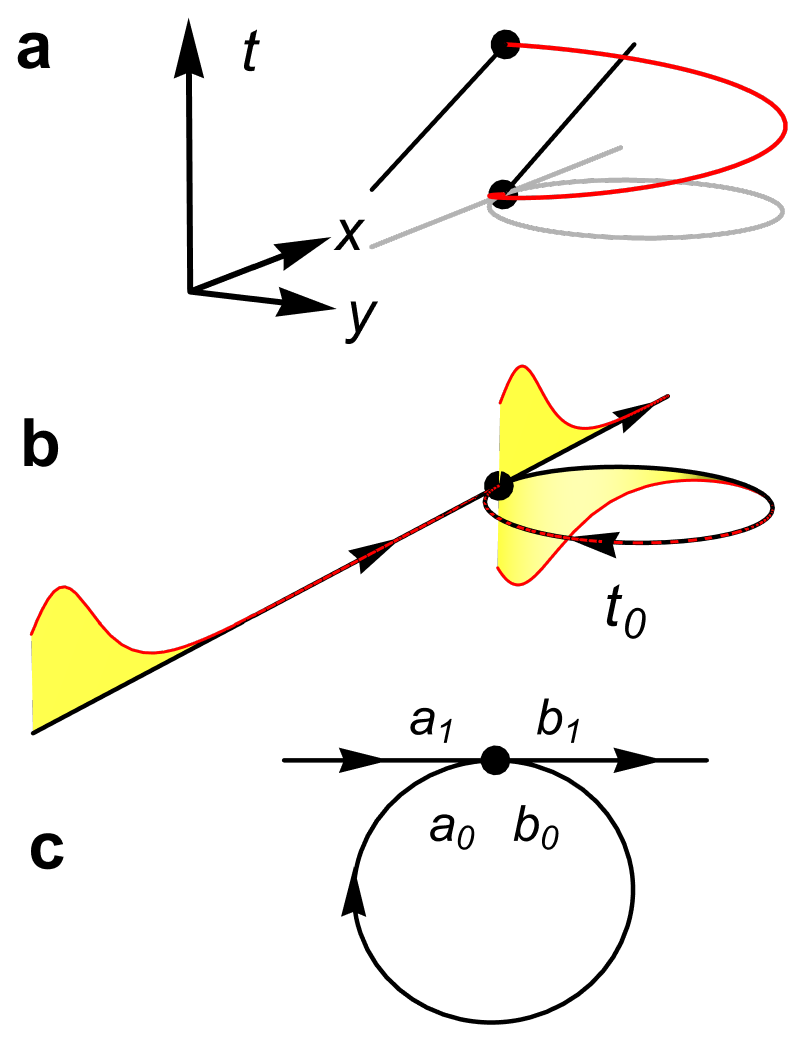}
		\caption{
			\small{Time travel. {\bf a}: Spacetime diagram of a time traveler. The particles of the traveler move at uniform velocity (black lines). When starting the ``time machine'' (an event marked by the upper dot) the traveler meets a precise copy of himself made of antiparticles and gets annihilated. The antiparticles (red) are made in the past (lower dot) together with the particles of the traveler who moves freely, while the antiparticles are on the way to meet his future self. The spatial trajectories are shown in gray. In {\bf b} --- the optical implementation of time travel --- the trajectories are replaced by a fibre loop of delay time $t_0$ connected via a parametric amplifier (dot) to a straight fibre; {\bf c} shows a top view and introduces the notation. In {\bf b} the amplifier is shown creating a pulse in the loop such that it  annihilates the incoming pulse when it arrives. As a result, the amplifier generates a partner pulse advanced in time by $t_0$. 			}
			\label{fig:traveling}}
	\end{center}
\end{figure}

Time travel has been a fascinating subject of fiction rather than science, but science can and does make statements whether one can travel in time and what the consequences would be, as we do in this paper:  we suggest a feasible experimental demonstration of time travel in quantum optics and work out its fundamental limitations. The background of our work is the following. The classical theory of space and time, general relativity, does allow time travel. Under certain circumstances, for example in G\"{o}del's universe \cite{Goedel}, travelers may return to an earlier coordinate time while advancing in their own proper time, drawing closed timelike curves in spacetime. Classical physics would allow time travel, but Hawking conjectured \cite{CPC} that quantum mechanics will prevent it. Closed spacetime curves would create quantum instabilities, making time travel inherently hazardous and ultimately impossible \cite{CPC}. Hawking's Chronology Protection Conjecture \cite{CPC} is not uncontested, though: Deutsch argued \cite{Deutsch} that time travel can be made consistent with quantum mechanics, and Deutsch's idea was demonstrated in a quantum simulation \cite{Simulation}. But this was a simulation in space, not in time. How would a real time traveler fare? Here we propose and analyze a simple optical analogue (Fig.~\ref{fig:traveling}) for testing Hawking's conjecture and for making  the quantum limits of time travel precise and quantitative. 

Our idea is related to temporal cloaking, also known as history editing \cite{McCall,Fridman,Lukens}, to analogues of gravity \cite{BLV,Unsch,Faccio,Jacquet,Barcelo},  time--varying optical materials \cite{Agrawal,Engheta,Mendonca,Galiffi} and to work on the notion of time in relativistic quantum information \cite{Zych1,Zych2,Procopio,Brukner}. It is inspired by the visualization \cite{Schleich} and simplification \cite{Mallary} of G\"{o}del's original idea \cite{Goedel} and by an intriguing electronic analogue of time travel \cite{Kitano}. Instead of an actual time traveler we consider an optical pulse going back in time. This would suffice for gaining information about events before they have happened, and it would avoid the hazards of material time travel.

To understand why time travel is potentially hazardous and how to implement it in optics, imagine a fictitious time traveler starting up a time machine for visiting the past (following the world line illustrated in Fig.~\ref{fig:traveling}a). The traveler consists of particles, and particles going back in time appear as antiparticles going forward in time. These antiparticles have been created in the past when the machine has stopped, together with the particles of the traveler who is then at liberty to leave the machine. For this process to be consistent, the created earlier self of the traveler --- and its antiparticle companion --- must be an exact replica of the traveler in the present. When starting the time machine, the traveler is instantly annihilated, facing certain { destruction} in the present and uncertain { restoration} in the past. 

It is therefore wise to resort to light pulses going back in time. The quantum particles of light --- the photons --- are their own antiparticles. and pair creation and annihilation are easily facilitated by optical parametric amplification \cite{OPA}. Our scheme (Fig.~\ref{fig:traveling}b) incorporates the features of time travel: light pulses travel in a straight fibre connected to a fibre loop by a parametric amplifier. In the loop, the pulses travel back into the past, and the amplifier provides the required annihilation and creation processes. Our scheme resembles Yariv's critical coupling resonator \cite{Yariv}, but with an active link \cite{Nemet} between fibre and loop instead of a passive coupling. It differs from the Hartmann effect \cite{Hartmann} in tunnelling \cite{Nimtz} as it preserves the shapes and amplitudes of light pulses. It is inspired by the electronic version of time travel \cite{Kitano} where electric pulses are connected to a resistive delay by an operational amplifier, a nonlinear device, but in our case the amplifier is linear and the scheme is, in--principle, perfectly faithful for classical light, which allows us to study the quantum limits of time travel. 

\section{Classical optics}

Let us first describe the system classically. In the fibre and in the loop, light simply propagates with uniform velocity $c$ (we include the refractive index in $c$). The critical point is the interaction between fibre and loop, and the delay in the loop. The time--dependent complex field amplitude $a_1(t)$ in the fibre infinitesimally before the point of contact  and the corresponding amplitude $a_0(t)$ in the loop are related to the amplitudes $b_1(t)$ and $b_0(t)$ immediately after the contact, by the well--known transformations \cite{OPA}:
\begin{align}
b_1(t) &= a_1(t) \cosh\zeta + a_0^*(t)\sinh\zeta \,,\quad
b_0(t) &= a_0(t) \cosh\zeta + a_1^*(t) \sinh\zeta 
\label{eq:rel}
\end{align}
where $\zeta$ parameterizes the gain $(\cosh\zeta)^2$ of the amplifier, assumed to be operating with undepleted pump \cite{OPA}. The amplitude $a_0$ of the light in the loop in front of the amplifier is given by the time--delayed amplitude $b_0$ after the amplifier as
\begin{equation}
a_0(t) = b_0(t-t_0)\,\cos\gamma
\label{eq:loop}
\end{equation}
where $t_0$ denotes the time delay in the loop and $\cos\gamma$ quantifies losses (in practice primarily coupling losses, $\cos\gamma>0$). To solve the system of Eqs.~(\ref{eq:rel}-\ref{eq:loop}) we use Fourier transformation. From Eq.~(\ref{eq:rel}) we get 
\begin{equation}
a_0^* = b_0^*\, \mathrm{sech}\,\zeta - a_1 \tanh\zeta \,,
\label{eq:a0}
\end{equation}
and the same for the Fourier transforms $\widetilde{a_0^*}$, $\widetilde{b_0^*}$ and $\widetilde{a_1}$. From Eq.~(\ref{eq:loop}) follows 
\begin{equation}
\widetilde{a_0^*} = \widetilde{b_0^*}\,e^{\mathrm{i}\omega t_0}\,\cos\gamma\,,
\end{equation}
and from Eq.~(\ref{eq:a0}) we obtain:
\begin{equation}
\widetilde{b_0^*} = \frac{\widetilde{a_1}\,\sinh\zeta}{1-e^{\mathrm{i}\omega t_0}\cosh\zeta \cos\gamma} \,.
\label{eq:b0fourier}
\end{equation}
Suppose the loss exceeds the gain, $\cosh\zeta \cos\gamma<1$. We expand the expression $(1-e^{\mathrm{i}\omega t_0}\cosh\zeta \cos\gamma)^{-1}$ in the geometric series $\sum_{m=0}^\infty (e^{\mathrm{i}\omega t_0}\cosh\zeta \cos\gamma)^m$, translate the Fourier transform $\widetilde{a_1}e^{\mathrm{i}\omega t_0}$ to $a_1(t-m t_0)$ to the time domain, and get the solution:
\begin{equation}
b_0^*(t) =  \sinh\zeta \sum_{m=0}^\infty (\cosh\zeta\cos\gamma)^m \, a_1(t-mt_0) \,.
\label{eq:b0loss}
\end{equation}
Finally, inverting Eqs.~(\ref{eq:rel}) gives the relation $a_1=b_1\cosh\zeta - b_0^*\sinh\zeta$ and hence
\begin{equation}
b_1(t)= a_1(t) \,\mathrm{sech}\,\zeta + b_0^*(t) \tanh\zeta \,.
\label{eq:b1}
\end{equation}
The output $b_1(t)$ depends only on $a_1(t-mt_0)$. The first term is given by the initially amplified signal $a_1(t)\cosh\zeta$ and is followed by a ringdown of pulses $a_1(t-mt_0)$ ($m>1$) with prefactors $(\sinh\zeta)^2 \cos\gamma (\cosh\zeta\cos\gamma)^{m-1}$. These factors comprise the incoupling and outcoupling coefficients $\sinh\zeta$, the coupling loss $\cos\gamma$ and the amplitude after $m-1$ cycles in the loop. Clearly, the initial pulse does not advance in time. 

\begin{figure}[h]
\begin{center}
\includegraphics[width=24pc]{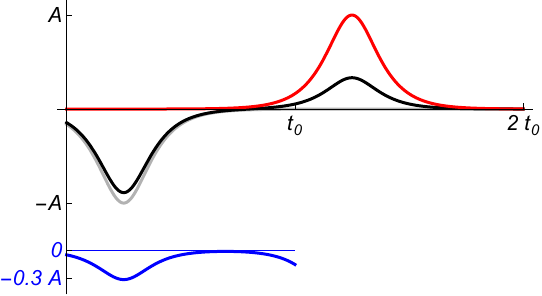}
\caption{
\small{Time travel with modest gain. Input pulse $a_1(t)$ [red curve] versus the output pulse $b_1(t)$ [black curve, Eqs.~(\ref{eq:b1}) and (\ref{eq:b0gain})] compared with $-a_1(t+t_0)$ [grey curve] for a gain of $10$ ($\cosh^2\zeta=10$) and no loss. Only the envelopes of the pulses are shown (for their optical oscillations see Fig.~\ref{fig:frames}). The separate plot in blue shows the initial light in the loop [Eqs.~(\ref{eq:loop}) and (\ref{eq:b0gain})]. As input pulse we took $a_1=A\,\mathrm{sech}[-10(t-1.25\,t_0)/t_0)]$ where the bulk of the incident pulse arrives in the interval $[t_0,2t_0]$. We clearly see that the output leaves before the input has arrived, and with good fidelity ($80\%$, see Appendix B) even for modest gain. 
}
\label{fig:pulses}}
\end{center}
\end{figure}

Now consider the case when the gain exceeds the loss, $\cosh\zeta\cos\gamma>1$. In this case, the previous geometric series does not converge, but we may divide both the numerator and the denominator of Eq.~(\ref{eq:b0fourier}) by $e^{\mathrm{i}\omega t_0}\cosh\zeta \cos\gamma$ and expand $(1-e^{-\mathrm{i}\omega t_0}\,\mathrm{sech}\,\zeta\,\mathrm{sec}\,\gamma)^{-1}$ as $\sum_{m=0}^\infty (e^{\mathrm{i}\omega t_0}\cosh\zeta\cos\gamma)^{-m}$. We thus obtain in the time domain:
\begin{equation}
b_0^*(t) =  -\sinh\zeta \sum_{m=1}^\infty (\cosh\zeta\cos\gamma)^{-m} \, a_1(t+mt_0)
\label{eq:b0gain}
\end{equation}
with Eq.~(\ref{eq:b1}) for $b_1$ as before. Note that Eqs.~(\ref{eq:b1}) and (\ref{eq:b0gain}) remain solutions of Eqs.~(\ref{eq:rel}) and (\ref{eq:loop}) for $t\ge t_\mathrm{amp}$ when the amplifier is switched on at $t_\mathrm{amp}$ (for example at $t_\mathrm{amp}=0$) but not acting before. Our solution shows that the incident light may advance in time: the pulse may leave the device before it has entered it {(Fig.~\ref{fig:pulses})}.

In the limit of strong amplification ($\zeta\rightarrow\infty$) we get $b_1(t)=-a_1(t+t_0)/\cos\gamma$. If there are no losses ($\gamma=0$) the incident pulse travels in time by the time $t_0$ set by the delay in the loop, without changing its shape and amplitude. The only effect of time travel is an overall sign change. With losses, the amplifier needs to enhance the signal by $(\cos\gamma)^{-1}$ such that, when delayed in the loop, it matches the incoming light for annihilating it. For finite amplification { (Fig.~\ref{fig:pulses})} the fidelity of the device is reduced, but whenever the gain exceeds the loss, time travel is possible. 

Yet there is one important caveat. Mathematically, our solution [Eq.~(\ref{eq:b0gain})] presumes that the Fourier transform of the outgoing light exists. Physically, this means that the pulse does not grow indefinitely while cycling in the loop and being amplified. To prevent this, the initial light in the loop must be launched such that it anticipates the pulse and can cancel it. The initial pulse in the loop may just annihilate the precursor of the pulse --- and not the pulse itself --- but it must anticipate it precisely. From Eq.~(\ref{eq:a0}) in the limit of large amplification we obtain $a_0^*\sim - a_1$. So even if the wing of an incoming pulse is exponentially small, the light in the loop must anticipate and match it. { This shows explicitly in our scheme (Fig.~\ref{fig:traveling}) that, while time travel is possible, it still cannot contradict causality\footnote[2]{Causality mplies the Kramers--Kronig relations \cite{KK} as the Fourier transforms of response functions are analytic on the upper half plane, but our amplifier introduces poles on the lower half.}. The past and the present must be absolutely consistent --- even in time travel.} 

The need for precise consistency for classical light already indicates that there will be problems for quantum light, as the amplitudes of quantum fields cannot be perfectly precise, but must fluctuate. This is a consequence of Heisenberg's uncertainty relation: the real and imaginary parts of the complex field amplitude behave like position and momentum, and both can never be precise at the same time.  

Another concern is this: the condition for time travel, gain exceeding loss, is the condition for lasing. The amplifier with feedback loop will act like a laser, producing light of its own from amplified vacuum fluctuations. In practice, the optical time machine should only run for a limited transient time greater than $t_0$ but sufficiently short for not drowning the signal in the background light. Note that the laser action of the machine corresponds { (Appendix A)} to the instability that led Hawking to the Chronology Protection Conjecture \cite{CPC}. Here we have reproduced this instability in its most elementary form.

\section{Quantum optics}

\begin{figure}
\begin{center}
\includegraphics[width=30pc]{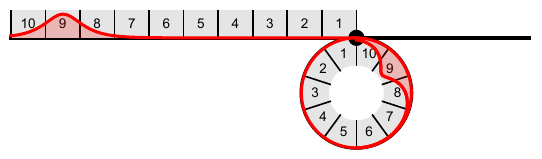}
\caption{
\small{Box modes. We quantize the field in terms of box modes (boxes, labeled by integers) that are long in comparison with the wavelength and short in comparison with the variations of the pulse amplitudes (red). The modes interact with each other at the amplifier (dot). 
}
\label{fig:boxes}}
\end{center}
\end{figure}

Let us work out the quantum--mechanical consequences. Quantum light is described by the operator $\widehat{A}$ of the vector--potential component corresponding to the relevant optical polarization. We may represent the quantum field of light in the mode expansion \cite{Book}:
\begin{equation}
\widehat{A} = \sum_k \left(\widehat{a}_k A_k + \widehat{a}_k^\dagger A_k^*\right) 
\label{eq:modex}
\end{equation}
where the $A_k$ are mode functions and the $\widehat{a}_k$ mode operators (labeled by $k$). The mode functions describe how the electromagnetic field of a single photon propagates in space $x$ and time $t$. For this, the $A_k$ must satisfy the classical wave equation and be normalized according to the scalar product \cite{Book}
\begin{equation}
(A_1,A_2) = \frac{\mathrm{i}}{\hbar} \int \left(A_1^*\partial_t A_2 - A_2 \partial_t A_1^* \right) \mathrm{d}x \,.
\end{equation}
If the mode functions satisfy the orthonormality conditions $(A_k,A_{k'})=\delta_{kk'}$ and $(A_k^*,A_{k'})=0$, the mode operators are guaranteed \cite{Book} to obey the Bose commutation relations $[\widehat{a}_k,\widehat{a}_{k'}^\dagger]=\delta_{kk'}$. Apart from those requirements, the mode functions can have any shape. For being able to describe fields with arbitrary amplitudes, we employ ``box modes'' (Fig.~\ref{fig:boxes}) that are short in comparison with the spatial amplitude variations and long in comparison with the wavelength. We sort the boxes into integer numbers $l$ and $m$. The $l$ describes the number of turns in the loop, while the $m$ subdivides the light field into $M$ boxes within one loop, assuming $M\gg 1$. At time $t$ the box mode in the fibre is located at $x=c(t - t_{l,m})$ with $t_{l,m}=(l+\frac{m}{M})t_0$ while in the loop the mode is located at $y=c\,\mathrm{mod}(t-t_{0,m},t_0)$. In both cases, $m$ runs from $0$ to $M-1$. 

The initial quantum state of light shall be the state closest to a perfect classical wave: a coherent state \cite{Book}. This is the most robust case for pure states, as nonclassical  states are more fragile \cite{Book}. For each box mode in the fibre, we assume the coherent state $|\alpha_{l,m}\rangle$ with $\alpha_{l,m}$ given by the classical complex amplitude at the position of the box,  {\it i.e.}\ by $a_1(t)$ at time $t=-t_{l,m}$. In the loop we assume for each mode the coherent state consistent with time travel: $|\alpha_m\rangle$ with $\alpha_m=a_0(-t_{0,m})$ and $a_0$ given by Eqs.~(\ref{eq:a0}) and (\ref{eq:b0gain}). We then switch the amplifier on and let it run for $L$ rounds in the loop. After passing the amplifier at the contact point ($x=0$) the operators of the fibre modes are transformed as
\begin{equation}
\widehat{a}_{1,l,m}' = \widehat{a}_{1,l,m}\cosh\zeta+\widehat{a}_{0,l,m}^\dagger \sinh\zeta\,,
\label{eq:trans1}
\end{equation}
while, after the $l$--th round in the loop, the mode operators turn into  
\begin{equation}
\widehat{a}_{0,l,m}' = \widehat{a}_{0,l,m}\cosh\zeta+\widehat{a}_{1,l,m}^\dagger \sinh\zeta\,,
\end{equation}
with the understanding 
\begin{equation}
\widehat{a}_{0,l+1,m} = \widehat{a}_{0,l,m}' \cos\gamma + \widehat{a}_{2,l,m} \sin\gamma\,.
\label{eq:trans3}
\end{equation}
Note that in Eq.~(\ref{eq:trans3}) --- describing loss --- we have introduced the mode operators $\widehat{a}_{2,l,m}$. The mathematical reason is the following: without the extra term the operators $\widehat{a}_{0,l+1,m}$ would no longer satisfy the Bose commutation relations, but with the $\widehat{a}_{2,l,m} \sin\gamma$ term they do. Only in this case, they are proper Bose operators and describe quantum light. Physically, the loss modes account for the additional quantum fluctuations absorption introduces according to the quantum version of the fluctuation--dissipation theorem \cite{Book}. 

Relations (\ref{eq:trans1}-\ref{eq:trans3}) (with $\gamma=0$) have appeared in the literature \cite{Corley,LP,Gaona} although in a different context and only theoretically so far \cite{Stein,Tettamanti,BHLComment,SteinCorrection}: they describe the action of a black--hole laser. The two mirrors of such a laser would be made by one black--hole and one white--hole horizon, confining radiation within. But these mirrors are also amplifiers, generating Hawking radiation \cite{Brout}. They may build up a macroscopic field due to cavity feedback. We see that our scheme (Fig.~\ref{fig:traveling}b) represents an optical analogue of black--hole lasing as well. Whether the device acts as a black--hole laser or a time machine depends on the initial conditions. For black--hole lasing, the initial field is in the vacuum state, for time travel it should be in the synchronized coherent state in loop and fibre according to Eqs.~(\ref{eq:a0}) and (\ref{eq:b0gain}). In this paper we focus exclusively on the case of time travel. 

Relations (\ref{eq:trans1}-\ref{eq:trans3}) describe the transformations of the mode operators in the Heisenberg picture. To capture the evolution of the quantum state in the Schr\"{o}\-dinger picture, we use Wigner functions \cite{Book,SchleichBook}. For this we write any of the $\widehat{a}$ operators in terms of the quadratures $\widehat{q}$ and $\widehat{p}$ as $\widehat{a}=2^{-1/2}(\widehat{q}+\mathrm{i}\widehat{p})$. The Bose commutation relations imply that $\widehat{q}$ and $\widehat{p}$ obey the commutation relations of position and momentum (with $\hbar=1$). The Wigner function $W(q,p)$ is a quasiprobability distribution \cite{Book} for those $q$ and $p$ giving intuition to quantum noise \cite{SchleichBook}. For example, the vacuum state has the Gaussian Wigner function $W=\pi^{-1}\exp(-q^2-p^2)$ while a coherent state with amplitude $\alpha=2^{-1/2}(q_0+p_0)$ corresponds to a Gaussian Wigner function with the same width as the vacuum noise, displaced by $q_0$ and $p_0$: $W=\pi^{-1}\exp[-(q-q_0)^2-(p-p_0)^2]$ behaving indeed like a classical probability distribution for $q$ and $p$ \cite{Book}. In addition to being intuitive, Wigner functions have another advantage: for mode transformations of the type (\ref{eq:trans1}-\ref{eq:trans3})  the Wigner function evolves like a classical probability distribution \cite{Simple} where the variables are transformed with the inverse of the transformations: 
\begin{align}
q_{1,l.m} &= q_{1,l,m}'\cosh\zeta-q_{0,l,m}' \sinh\zeta \,,
\nonumber\\
q_{0,l,m} &= q_{0,l,m}\cosh\zeta-q_{1,l,m}' \sinh\zeta \,,\\
q_{0,l+1,m}' &= q_{0,l,m}\cos\gamma - q_{2,l,m}' \sin\gamma\,, \nonumber
\label{eq:qtrans}
\end{align}
and the same for the $p_{\nu,l,m}$ with $\zeta$ replaced by $-\zeta$. From this follows that the amplitude of an initial coherent state evolves exactly like the classical amplitude, while the amplitude noise is the amplified vacuum noise, the optical analogue of black--hole lasing \cite{Corley,LP,Gaona}. 

For predicting the extra quantum noise produced by time travel, it is therefore sufficient to consider the apparatus with vacuum input. The reduced quantum state for each box mode must be a state with a Gaussian Wigner function. As the variances of the transformed quadratures, evaluated in the vacuum state, are even functions of $\zeta$ this Gaussian must be isotropic in $q$ and $p$: it is a thermal state \cite{Book}. To quantify the thermal noise we calculate the particle number in the fibre produced for vacuum input after $L$ rounds in the loop:
\begin{equation}
\langle 0 |\, \widehat{a}'^\dagger_{1,l,m}\widehat{a}'_{1,l,m}\,|0\rangle = (\sinh\zeta)^2(\cosh\zeta\cos\gamma)^{2L} = \bar{n}\,.
\label{eq:noise}
\end{equation}
In our case, this particle number quantifies the extra quantum noise produced by time travel for the most favourable initial condition, an incident coherent state. { The time--traveled quantum state is a displaced thermal state \cite{Book} described by the Wigner function $W=(s/\pi)\exp[-s(q-q_1)^2-s(p-p_1)^2]$ with $s=(1+2\bar{n})^{-1}$ and the quadratures $q_1$ and $p_1$ of the classical solution.} Small classical deviations from the required initial conditions are similarly amplified. While the coherent amplitude of a light pulse would advance in time, the quantum noise grows exponentially as quantified by the characteristic particle number of Eq.~(\ref{eq:noise}). Therefore, while the time travel of light would be possible in classical optics, quantum optics imposes severe limitations. 

\section{Conclusions}

{ Our proposed optical experiment illuminates several aspects of time travel. On the one hand, an optical pulse may travel in time --- in the sense that the pulse leaves a device before having entered it (Fig.~\ref{fig:pulses}). On the other hand, an initial partner pulse also needs to be launched (Fig.~\ref{fig:pulses} in blue) that precisely matches the incoming pulse about to travel in time ---  the past and the present need to be consistent. Classical time travel is possible, but in quantum physics Heisenberg's uncertainty relation implies that the amplitude of light fields cannot be perfectly precise, which limits the precision of the initial light and causes quantum limits to time travel. They appear as additional thermal noise, spoiling the quantum fidelity of the ``time machine''. All this resonates with} Hawking's Chronology Protection Conjecture \cite{CPC}. Hawking writes \cite{CPC}: ``If one tries to warp spacetime to allow travel into the past, vacuum polarization will cause the energy-momentum tensor to increase.''  We have derived the optical equivalent of this statement: the optical time machine acts like a laser, it amplifies vacuum noise by stimulated emission of radiation. Then Hawking states the obvious: ``There is also strong experimental evidence in favor of the conjecture from the fact that we have not been invaded by hordes of tourists from the future.'' However, our analysis suggests that the picture is not as black and white: while perfect time travel is forbidden, imperfect time travel is allowed. If the optical equivalent (Fig.~\ref{fig:traveling}) of the required warped space--time  is switched on for a finite period, the extra quantum noise of time travel is finite and possibly manageable. We have suggested a perfectly feasible experiment to test these quantum limits of time travel in the laboratory. 

\section*{Acknowledgments} 

We thank
I. Gabai,
T. Kiss,
R. Lapkiewicz,
N. Nemet,
L. M. Procopio,
A. Rauschenbeutel,
Ph. Schneeweiss,
I. Smith, 
and
J. Volz
for illuminating discussions. 
Our work was supported by the Murray B. Koffler Professorial Chair, a Global Fellowship of the Vienna Institute of Technology  (U.L.) and by the Marcos Moshinsky Chair (D.B.).

\newpage

\appendix

\section*{Appendix}

\section{Hawking's Chronology Protection Conjecture}

In this appendix we show that our optical scheme directly implements the key argument of Hawking's Chronology Protection Conjecture \cite{CPC}. To see this, we translate Hawking's paper into optics. Hawking shows \cite{CPC} that time travel in general relativity necessitates the existence of closed null geodesics (or geodesics infinitely close to closed null geodesics). Null geodesics are the space--time trajectories of light rays as in our case. Hawking then argues that, after each round on such a closed null geodesic, the tangent vector grows (or remains the same in pathological cases) but never shrinks. Now, the tangent vector corresponds to the four vector $(E,p^ic)$ of energy and momentum with  $E=pc$ in a local geodesic frame (as it is a null vector). So, if the tangent vector grows after each round, energy and momentum must grow, which is the defining feature of an amplifier. Hawking thus arrives at the same physical system (Fig.~\ref{fig:traveling}) as in our case, except that our amplifier is localized and can be switched on and off, whereas in a space--time loop amplification takes place continuously and indefinitely.  Hawking finds that the two--point quantum correlation function diverges, which would generate a strong gravitational field counteracting the time travel. In our case, the amplifier ring would deplete its energy source if it were to run indefinitely. In both cases, time travel will be related to lasing, and the divergence Hawking has worked out corresponds to the amplified vacuum noise of a laser. Not every laser is a time machine, but as Hawking's geometrical argument suggests, every time machine is a laser. 

\newpage

\section{Possible experimental implementation}

\begin{figure}[h]
\begin{center}
\includegraphics[width=20pc]{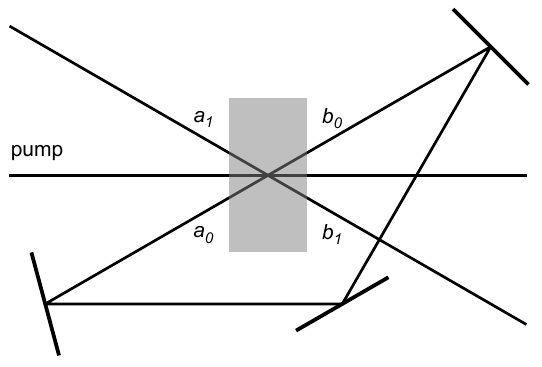}
\caption{
\small{Setup. Our scheme (Fig.~\ref{fig:traveling}) can be implemented in free--space optics using a $\chi^{(2)}$ crystal \cite{Boyd} as parametric amplifier \cite{OPA} (grey box). The signal mode of the amplifier serves as the signal in our setup, with amplitudes $a_1$ before and $b_1$ after amplification. The idler \cite{OPA}, with amplitudes $a_0$ before and $b_0$ after amplification, is reflected by mirrors to form the loop of the optical time machine (Fig.~\ref{fig:traveling}).
}
\label{fig:setup}}
\end{center}
\end{figure}

Although we have conceived our optical time machine as a fibre device (Fig.~\ref{fig:traveling}) it is perhaps easiest to implement in free--space optics using a $\chi^{(2)}$ crystal \cite{Boyd} for the required parametric amplification (Fig.~\ref{fig:setup}). { In this set--up losses can be kept extremely low. The two input and the two output beams are called signal and idler \cite{Boyd}. Their angles are determined by phase--matching conditions and the linear birefringence of the crystal \cite{Boyd}.}  The crystal is pumped by a laser beam with twice the carrier frequency of the time--traveling light. The pump beam can be switched on and off for the time the machine should work. 

\begin{figure}
\begin{center}
\includegraphics[width=28pc]{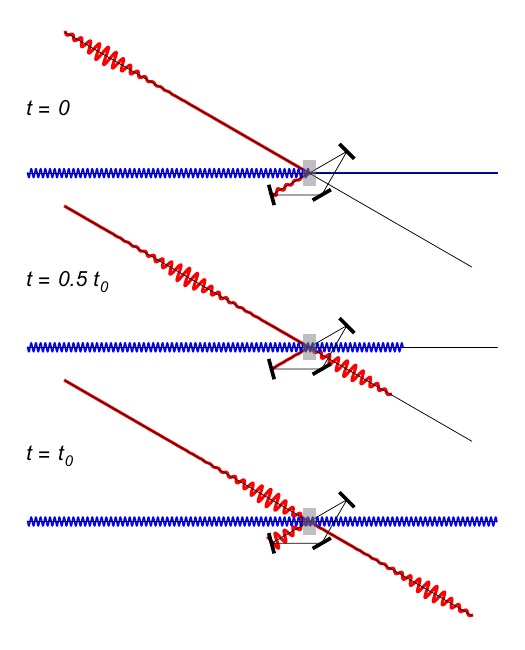}
\caption{
\small{Time frames { (for the pulses of Fig.~\ref{fig:pulses})}. The figure shows three typical phases in the process of the signal (red wave) traveling in time through the apparatus (Fig.~\ref{fig:setup}) pumped by a beam (blue wave) of twice the carrier frequency of the signal. In the initial phase ($t=0$) the pump just reaches the ampiifier crystal while the signal is still more than the delay time $t_0$ away. The delay line is filled with initial light [Eq.~(\ref{eq:b0gain})] that is about to re--create the precursor of the pump pulse after the crystal. Halfway through the process ($t=t_0/2$) the signal is created while the incident pulse has not reached the device yet. Finally ($t=t_0$) the pulse reaches the crystal and is about to be annihilated by the light built up in the delay line, completing the time travel. 
}
\label{fig:frames}}
\end{center}
\end{figure}

Figure~\ref{fig:frames} shows a sequence of time frames for the setup at different stages in the time--traveling process. Let $x$ be the coordinates of the incident light and $y$ the coordinate in the loop. The light propagates as $a_1(t-x/c)$ and $b_0(t-y/c)$. Initially ($t=0$) the pulse $a_1(-x)$ is incident at a greater distance than $ct_0$ away and the loop is filled with light matching the precursor of the pulse as $b_0(-y/c)$ for $y\in [0,t_0]$ and $b_0$ given by Eq.~(\ref{eq:b0gain}). The filling could be done {\it e.g.} by sending light through one of the mirrors in the loop (Fig.~\ref{fig:setup}) being made { slightly} semitransparent. The next panel shows the setup halfway through the time loop ($t=t_0/2$). One sees that an output pulse is created before the input pulse arrives. Finally ($t=t_0$) the incident pulse is being annihilated by the light inside the loop (playing the role of the ``anti--traveler'' in Fig.~\ref{fig:traveling}a) while the output pulse travels on. Note that only a modest gain of $\cosh^2\zeta=10$ is assumed for creating the light fields shown ({ Figs.~\ref{fig:pulses} and} \ref{fig:frames}). Nevertheless, the classical fidelity of time traveling, the overlap of the output pulse with the input pulse, is quite high. { With the standard definitition
$$
F = \left|\frac{\int_0^{t_0} a_1(t+t_0)^*b_1(t)\,\mathrm{d} t}{\int_{t_0}^{2t_0}a_1^*(t)a_1(t)\,\mathrm{d} t}\right|^2
$$
we get $F=0.81$ for the pulse shown (Figs.~\ref{fig:pulses} and \ref{fig:frames}) and a gain of 10. For larger gain the classical fidelity becomes virtually perfect.} 

Quantum--mechanically, the amplified spontaneous emission in the loop is going to create extra noise on the output pulse that can be measured by {\it e.g.} homodyne detection \cite{Book} with a copy of the initial pulse as local oscillator \cite{Book} (reference beam). { Homodyne detection can reach nearly perfect accuracy as the electronic noise of highly efficient photodiodes is made irrelevant by mixing with a high--power local oscillator \cite{Book}.} For a delay line in the order of $0.3\,\mathrm{m}$  (one foot)  $t_0\approx 1\,\mathrm{ns}$ { and so the typical pulse duration would be in the order of $1\,\mathrm{ns}$. Such pulses can be crafted from continuous--wave laser  light by traveling--wave electro--optic modulators \cite{Modulators} that routinely operate in the required GHz range. The pulse forms can be dialed--in electronically and precisely timed and controlled.}  The components of the experimental setup (Fig.~\ref{fig:setup})  are all standard equipment; { they only need to be put together with sufficient precision and skill.}


\newpage
\begin{thebibliography}{99}

\bibitem{Goedel}
K. G\"{o}del, Rev. Mod. Phys. {\bf 21}, 447 (1949).

\bibitem{CPC}
S. W. Hawking, Phys. Rev. D {\bf 46}, 603 (1992).

\bibitem{Deutsch}
D. Deutsch, Phys. Rev. D {\bf 44}, 3197 (1991).

\bibitem{Simulation}
M. Ringbauer, M. A. Broome, C. R. Myers, A. G. White, and T. C. Ralph, 
Nat. Commun. {\bf 5}, 4145 (2014).

\bibitem{McCall}
M. W. McCall, A. Favaro, P. Kinsler and A. Boardman,  
J. Opt. 13, 024003 (2011).

\bibitem{Fridman}
M. Fridman, A. Farsi, Y. Okawachi, and A. L. Gaeta, 
Nature {\bf 481}, 62 (2012).

\bibitem{Lukens}
J. M. Lukens, D. E. Leaird, and A. M. Weiner, 
Nature {\bf 498}, 205 (2013). 

\bibitem{BLV}
C.  Barcelo, S. Liberati, and M. Visser,
Living Rev. Relativity {\bf 8}, 12 (2005).

\bibitem{Unsch}
W. G. Unruh and R. Sch\"utzhold (eds.)
{\it Quantum Analogues: From Phase Transitions to Black Holes and Cosmology}
(Springer, Berlin, 2007).

\bibitem{Faccio}
D. Faccio, F. Belgiorno, S. Cacciatori, V. Gorini, S. Liberati, and U. Moschella (eds.),
{\it Analogue Gravity Phenomenology: Analogue Spacetimes and Horizons, from Theory to Experiment},
Lecture Notes in Physics {\bf 870}
(Springer, Cham, 2013).

\bibitem{Jacquet}
M. J. Jacquet, S. Weinfurtner and F. K\"{o}nig,
Phil. Trans. Roy. Soc. A  {\bf 378}, 20190239 (2020).

\bibitem{Barcelo}
C. Barcelo, J. Eguia Sanchez, G. Garcia-Moreno, and G. Jannes,
Eur. Phys. J. C {\bf 82}, 299 (2022).

\bibitem{Agrawal}
G. P. Agrawal, 
Photonics {\bf 12}, 611 (2025).

\bibitem{Engheta}
N. Engheta, 
Science {\bf 379}, 1190 (2024).

\bibitem{Mendonca}
J. T. Mendonca, Symmetry {\bf 16}, 1548 (2024).

\bibitem{Galiffi}
E. Galiffi, R. Tirole, S. Yin, H. Li, S. Vezzoli, P. A. Huidobro, M. G. Silveirinha, R. Sapienza, A. Alu, and J. B. Pendry,
Adv Photonics {\bf 4}, 014002 (2022).

\bibitem{Zych1}
M. Zych,
{\it Quantum systems under gravitational time dilation},
(Springer, Cham, 2017).

\bibitem{Zych2}
M. Zych, F. Costa, I. Pikovski, and C. Brukner, 
Nat. Commun. {\bf 2}, 505 (2011).

\bibitem{Procopio}
L. M. Procopio, A. Moqanaki, M. Araújo, F. Costa, I. Alonso Calafell, E. G. Dowd, D. R. Hamel, L. A. Rozema, Č. Brukner, Ph. Walther,
Nat. Commun.  {\bf 6}, 7913 (2015).

\bibitem{Brukner}
E. Castro-Ruiz, F.  Giacomini, A. Belenchia, and C. Brukner,
Nat. Commun.  {\bf 11}, 2672 (2020).

\bibitem{Schleich}
M. Buser, E. Kajari, and W. P. Schleich,  
New. J. Phys. {\bf 15}, 013063 (2013).

\bibitem{Mallary}
C. Mallary, G. Khanna, and R. H Price,  
Class. Quant. Grav. {\bf 35}, 175020 (2018).

\bibitem{Kitano}
M. Kitano, T. Nakanishi, and K. Sugiyama, 
 IEEE J. Sel. Top. Quant. Electron. {\bf 9}, 43 (2003).

\bibitem{OPA}
R. Paschotta, {\it Optical parametric amplifiers} in RP Photonics Encyclopedia.

\bibitem{Yariv}
A. Yariv,
Electron. Lett. {\bf 36}, 321 (2000).

\bibitem{Hartmann}
T. E. Hartmann,
J. Appl. Phys. {\bf 33}, 3427 (1962).

\bibitem{Nimtz}
A. Enders and G. Nimtz,
J. Phys. France {\bf 2}, 1693 (1992). 

\bibitem{Nemet}
N. Nemet and S. Parkins,
Phys. Rev. A {\bf 94}, 023809 (2016).

\bibitem{KK}
R. Paschotta, {\it Kramers–Kronig Relations} in RP Photonics Encyclopedia.

\bibitem{Book}
U. Leonhardt,
{\it Essential Quantum Optics} 
(Cambridge University Press, Cambridge, 2010). 

\bibitem{Corley}
S. Corley and T. Jacobson,  
Phys. Rev. D {\bf 59}, 124011 (1999).

\bibitem{LP}
U. Leonhardt and T. G. Philbin, {\it Black hole lasers revisited},
in Lect. Notes Phys. {\bf 718}, 229 (2007).

\bibitem{Gaona}
J. L. Gaona-Reyes and D. Bermudez,
Ann. Phys. (New York) {\bf 380}, 41 (2017). 

\bibitem{Stein}
J. Steinhauer, Nat. Phys. {\bf 10}, 864 (2014).

\bibitem{Tettamanti}
M. Tettamanti, S. L. Cacciatori, A. Parola, and I. Carusotto,  
Europhys. Lett. {\bf 114}, 60011 (2016).

\bibitem{BHLComment}
Y.-H. Wang, T. Jacobson, M. Edwards, and Ch. W. Clark, SciPost Phys. {\bf 3}, 022 (2017).

\bibitem{SteinCorrection}
V. I. Kolobov, K. Golubkov, J. R. de Nova, and J. Steinhauer, Nat. Phys. {\bf 17}, 362 (2021).

\bibitem{Brout}
R. Brout, S. Massar, R. Parentani, and Ph. Spindel, Phys. Rept. {\bf 260}, 329-446 (1995).

\bibitem{SchleichBook}
W. P. Schleich,
{\it Quantum optics in phase space}
(Wiley-VCH, Berlin, 2001).

\bibitem{Simple}
U. Leonhardt, 
Rep. Prog. Phys. {\bf 66}, 1207 (2003).

\bibitem{Boyd}
R. W. Boyd,
{\it Nonlinear optics} 
(Elsevier, Amsterdam, 2020).

\bibitem{Modulators}
R. Paschotta, {\it Electro-optic Modulators} in RP Photonics Encyclopedia.


\end{thebibliography}
\end{document}